\documentclass[11pt]{article}
\usepackage{moriond}

\usepackage{amsmath,amssymb}
\usepackage{array}

\def\be{\begin{equation}}
\def\ee{\end{equation}}
\def\bea{\begin{eqnarray}}
\def\eea{\end{eqnarray}}

\def\br{{\rm BR}}
\def\hsm{h_{\rm SM}}
\def\beq{\begin{equation}}
\def\eeq{\end{equation}}
\def\tanb{\tan\beta}

\def\gam{\gamma}

\def\gev{~{\rm GeV}}

\def\lsim{\mathrel{\raise.3ex\hbox{$<$\kern-.75em\lower1ex\hbox{$\sim$}}}}

\newcommand{\Photo}{\includegraphics[height=35mm]{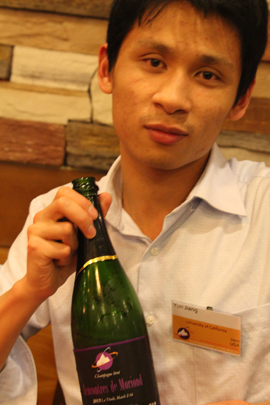}}

\begin{document}
\vspace*{-0.5cm}
\title{125~GeV HIGGS BOSONS IN TWO-HIGGS-DOUBLET MODELS~\footnote{This talk originates from the published paper~\cite{Drozd:2012vf} and will be contributed to proceedings of the conference "Rencontres de Moriond EW 2013", La Thuile, Italy, 2-9 Mar. 2013.}}

\author{ Yun JIANG}

\address{Department of Physics, University of California, Davis, CA 95616, USA}

\maketitle\abstracts{
Moriond 2013 ALTAS data at 125 GeV state appears to exhibit a substantial excess in the di-photon final state and in the $ZZ$ decaying to four lepton channel, whereas which are more or less SM-like rate observed by CMS MVA analysis. We examine the maximum Higgs signal enhancements that can be achieved in the 2HDM when either a single Higgs or more than one Higgs have mass(es) near 125 GeV. In general, the constraints of vacuum stability, unitarity and perturbativity play the key role in restricting possibilities of signal enhancement. The Type II model allows for an enhancement in the di-photon rate (relative to the SM) of the order of 2-3 but associated with an even larger $ZZ$  or too large $\tau\tau$ signal. In contrast, the maximal value of the di-photon signal in the Type I model can reach the order of 1.3 for which the ZZ signal is of order 1.}

\section{125~GeV Higgs signal at the LHC}

After over thirty years of waiting, on July 4th 2012 the Large Hadron Collider (LHC) and also the Tevatron discovered a Higgs-like resonance with mass at 125-126~GeV~\cite{Aad:2012tfa}~\cite{Chatrchyan:2012ufa}~\cite{Aaltonen:2012qt}. This signal was further confirmed up to $7\sigma$ significance at Council 2012~\cite{ATLAS:2012klq} and Moriond 2013~\cite{Chatrchyan:2013lba}. However, there are discrepancies between the two LHC collaborations in the gluon fusion induced $\gamma\gamma$, $ZZ^*\to 4\ell$ and $WW^*\to 2\ell 2\nu$ rates. The updated ATLAS data still shows substantial enhancement~\cite{ATLAS:2013oma}~\cite{ATLAS:2013nma} while the CMS MVA analysis finds them roughly SM-like~\cite{CMS:ril}~\cite{CMS:xwa}. At the moment one cannot draw the conclusion that it is the Standard Model (SM) Higgs boson although it is consistent with the SM value at 95 \% CL. The deviations of the signal rates, even very small, relative to the SM predictions provide valuable hints in constructing the nature of underlying theory and drive us to explore a more complicated model beyond the SM.

\section{Two-Higgs-doublet model}

Beyond the SM, two-Higgs-doublet model (2HDM) is the simplest extension. It employs a second Higgs doublet with the same hypercharge $Y=1$. The general 2HDM Higgs sector potential is 
\begin{equation}
  \begin{aligned}
    \mathcal{V} = &m_{11}^2\Phi_1^\dagger\Phi_1+m_{22}^2\Phi_2^\dagger\Phi_2
    -\left[m_{12}^2\Phi_1^\dagger\Phi_2+\mathrm{h.c.}\right]
    \\
    &+\frac{1}{2}\lambda_1\left(\Phi_1^\dagger\Phi_1\right)^2
    +\frac{1}{2}\lambda_2\left(\Phi_2^\dagger\Phi_2\right)^2
    +\lambda_3\left(\Phi_1^\dagger\Phi_1\right)\left(\Phi_2^\dagger\Phi_2\right)
    +\lambda_4\left(\Phi_1^\dagger\Phi_2\right)\left(\Phi_2^\dagger\Phi_1\right)
    \\&+\left\{
    \frac{1}{2}\lambda_5\left(\Phi_1^\dagger\Phi_2\right)^2
    +\left[\lambda_6\left(\Phi_1^\dagger\Phi_1\right)
      +\lambda_7\left(\Phi_2^\dagger\Phi_2\right)
      \right]\left(\Phi_1^\dagger\Phi_2\right)
    +\mathrm{h.c.}\right\}.
  \end{aligned}
  \label{eq:pot_gen}
\end{equation}
Assuming a real $m^2_{12}$ and $\lambda_6$, $\lambda_7$ to be zero, we study the $\mathcal{CP}$-conserved 2HDM with soft breaking of $Z_2$ symmetry ($\Phi_1\to \Phi_1$, $\Phi_2\to -\Phi_2$). This 2HDM structure generally contains two scalar Higgs bosons $h$, $H$ with mass ordering $m_h<m_H$, one pseudoscalar Higgs boson $A$ and one charged Higgs boson $H^\pm$. We choose to use the ``physical basis" in which the inputs are the physical Higgs masses ($m_h,m_H,m_A,m_{H^\pm}$), the  vacuum expectation value ratio ($\tan\beta$), and the $\mathcal{CP}$-even Higgs mixing angle, $\alpha$, supplemented by $m^2_{12}$. In this paper we discuss the Type I and the Type II models, that are distinguished by the fermion coupling pattern. In the Type I model all fermions couple to just one of the Higgs doublets while in the Type II model up-type fermions couple to one of the Higgs doublet while down-type fermions couple to the other one. In both two Types, dangerous tree level flavour-changing neutral current (FCNC) is completely absent.

Theoretically, vacuum stability, unitarity and coupling-constant perturbativity (denoted jointly as SUP) are required to be satisfied. Regarding the experiment constraints, we consider electroweak precision measurement $S,T,U$, LEP data and $B$ physics. The constraint from LEP essentially select the allowed parameter region in the $(\alpha, \beta)$ space and the one from $B$ physics set up a lower bound on charged Higgs mass with respect to $\tan\beta$. In addition, we ignore the anomalous magnetic moment of the muon $(g-2)_\mu$ as its 2HDM contribution is very small unless $\tan\beta$ is of order 100.

\section{125~GeV Higgs phenomenology in the 2HDMs}

We first examine the maximal gluon fusion ($Y=gg$) induced diphoton ($X=\gam\gam$) rate for certain 2HDM Higgs boson(s) $h_i$ defined as follows:
\beq
R^{h_i}_{Y}(X)\equiv {\sigma (Y\to h_i) \br(h_i\to X)\over \sigma (Y\to \hsm) \br(\hsm\to X)},
\eeq
that can be achieved in Type I and Type II 2HDM after imposing various combinations of the constraints outlined in the last section. Note that when considering cases where more than one Higgs has mass of $\sim 125\gev$~\cite{Gunion:2012gc}, we sum the different $R^{h_i}$ ($i=1$ for $h$, $i=2$ for $H$ and $i=3$ for $A$) for the production/decay channel of interest. Under various scenarios discussed later, whether Type I or Type II, whether degeneracy present or not, we observe in the figures below that for most values of $\tanb$ the $B$/LEP and STU precision electroweak constraints, both individually and in combination, leave the maximum $R^h_{gg}(\gam\gam)$ unchanged relative to a full scan over all input parameters (as specified earlier) prior to imposing any constraints. In contrast, the SUP constraints greatly reduce the maximum value of $R^h_{gg}(\gam\gam)$ that can be achieved and that value is left unchanged when $B$/LEP and STU constraints are imposed in addition.

\subsection{Single Higgs scenario: $m_h=125\gev$ or $m_H=125\gev$}\label{singleHiggs}

\vspace{-.2in}
\begin{figure}[h]
\begin{center}
\includegraphics[width=0.45\textwidth]{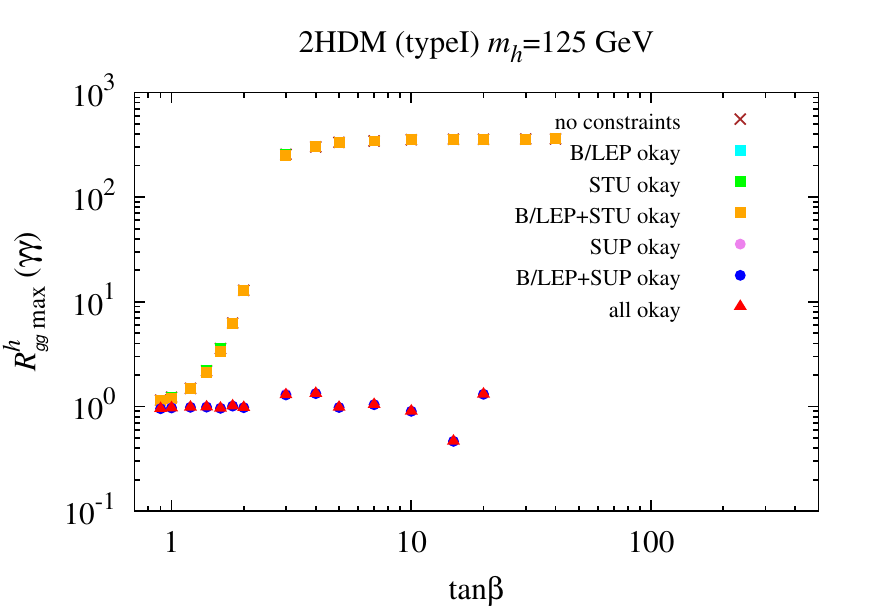}
\includegraphics[width=0.45\textwidth]{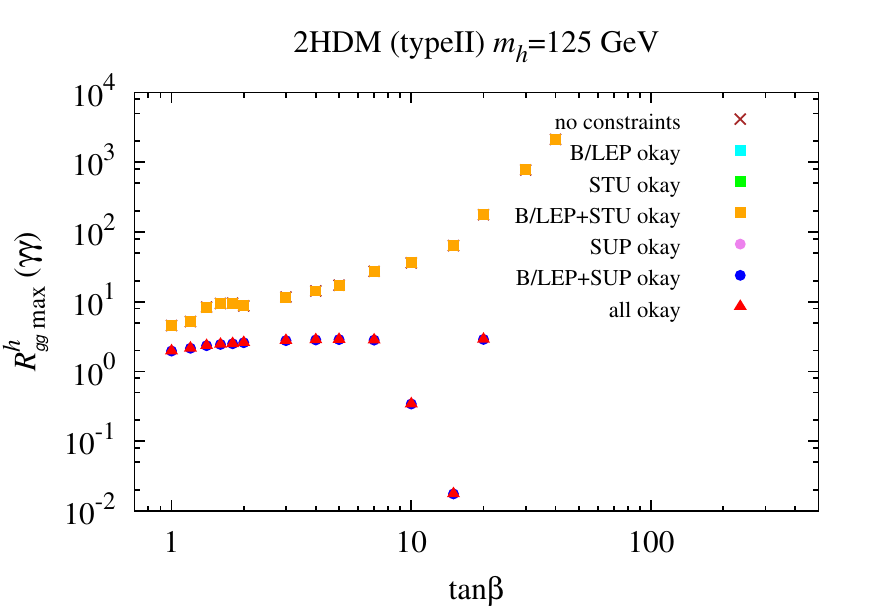}
\end{center}
\vspace{-.2in}
\caption{The maximum $R^h_{gg}(\gam\gam)$ values in the Type I (left) and Type II (right) models for $m_h=125\gev$ as a function of $\tanb$ after imposing various constraints --- see figure legend. }
\label{honly}
\end{figure}

As shown in Fig.~\ref{honly}, in the Type I model maximum $R^h_{gg}(\gam\gam)$ values highly above $1.3$ are not possible, with values close to $1$ being more typical for most $\tanb$ values. The maximal $R^h_{gg}(\gam\gam)$ is  of  order of $1.3$, as found if $\tanb=4$ or  $20$. In these cases,  $R^h_{gg}(ZZ)$ and $R^h_{gg}(\tau\tau)$ are of order 1, together with light charged Higgs $m_{H^\pm}=90\gev$. In contrast,  maximum $R^h_{gg}(\gam\gam) $ values in the range of $2-3$ are possible for $2 \leq\tanb\leq 7$ and $\tanb=20$ in the Type II model. However, the value of $R^h_{gg}(ZZ)$ corresponding to the parameters  that maximize $R^h_{gg}(\gam\gam)$ is typically large, $\sim 3$. In fact, $R^h_{gg}(ZZ)> R^h_{gg}(\gam\gam)$ whenever $R^h_{gg}(\gam\gam)$ is even modestly enhanced. The current experimental situation is confused. 
The Moriond 2013 ATLAS data shows  central values of $R^h_{gg}(\gam\gam)\sim 1.6$ and $R^h_{gg}(ZZ)\sim 1.5$~\cite{ATLAS:2013oma}~\cite{ATLAS:2013nma}. In the Type II model case, the former would imply $R^h_{gg}(ZZ)> 2$, somewhat inconsistent with the observed central value.  However, the data uncertainties are significant and so it is too early to conclude that the Type II model cannot describe the ATLAS data. The Moriond  2013 CMS data has central values of  $R^h_{gg}(\gam\gam)<1$ and $R^h_{gg}(ZZ)\sim 1$~\cite{CMS:ril}~\cite{CMS:xwa}, a situation completely consistent with the Type II model predictions.

\begin{figure}[h]
\begin{center}
\includegraphics[width=0.45\textwidth]{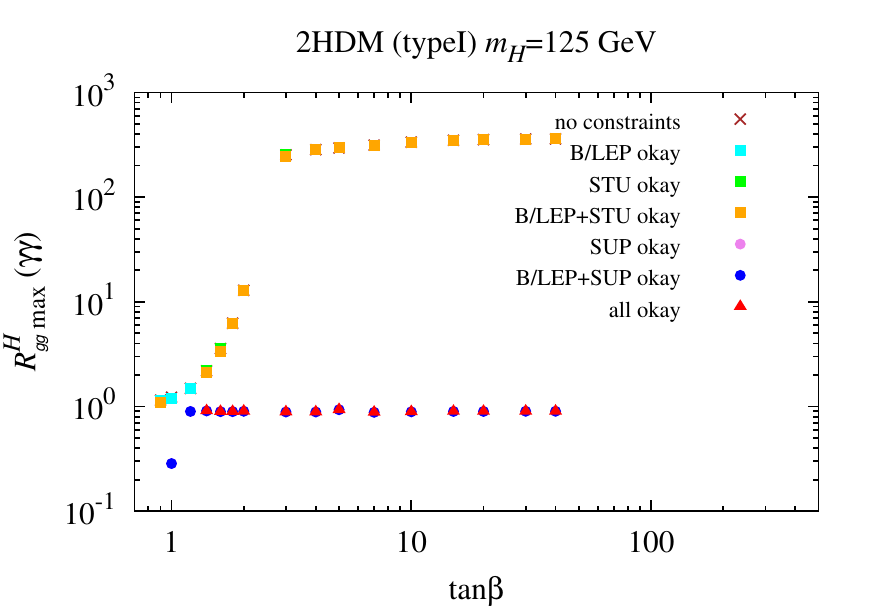}
\includegraphics[width=0.45\textwidth]{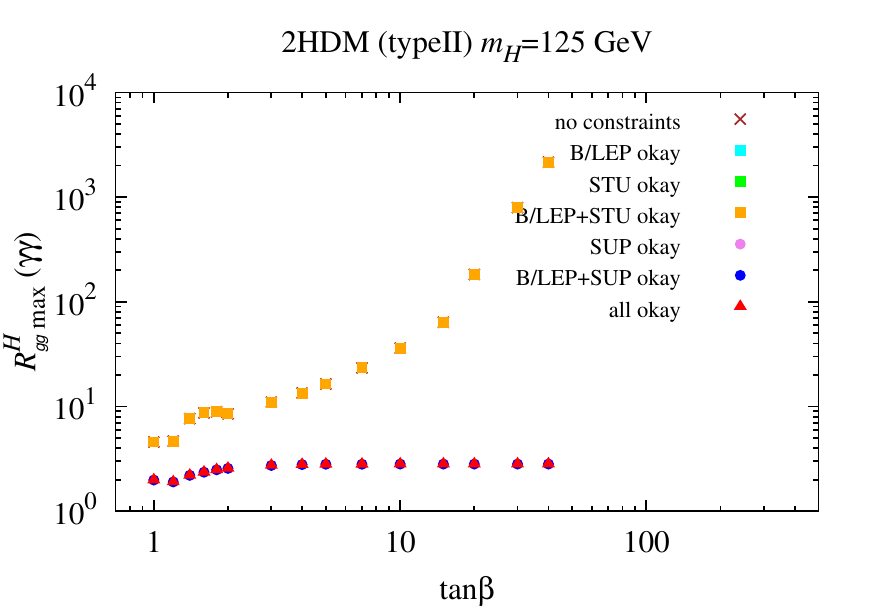}
\end{center}
\vspace{-.2in}
\caption{The maximum $R^H_{gg}(\gam\gam)$ values for $m_H=125\gev$ w.r.t. $\tanb$ after imposing various constraints. }
\label{Honly}
\end{figure}

Corresponding results for the $H$ are presented for the Type I and Type II models in Fig.~\ref{Honly}.  In the case of the Type I model, an enhanced gluon fusion induced $\gam\gam$ rate does not seem to be possible after imposing the SUP constraints, whereas maximal enhancements of order $R^H_{gg}(\gam\gam)\sim 2.8$ are quite typical for the Type II model, albeit with even larger $R^H_{gg}(ZZ)$.  Again, in the case of the Type II model $R^H_{gg}(\gam\gam)<R^H_{gg}(ZZ)$ applies more generally whenever $R^H_{gg}(\gam\gam)$ is significantly enhanced.

\subsection{Degenerate Higgs scenario: $m_h=125\gev$ and $m_A \sim 125\gev$}\label{degHiggs}

\begin{figure}[h]
\begin{center}
\includegraphics[width=0.45\textwidth]{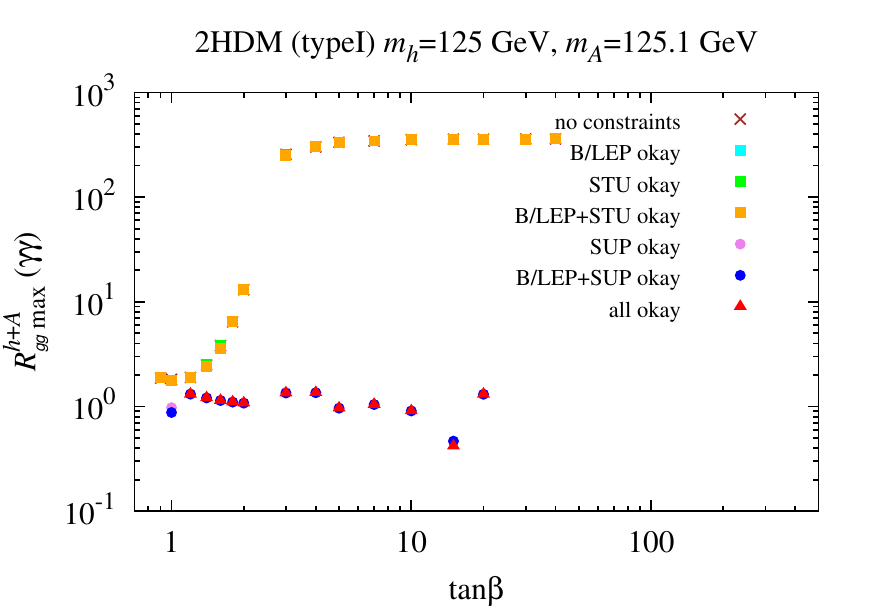}
\includegraphics[width=0.45\textwidth]{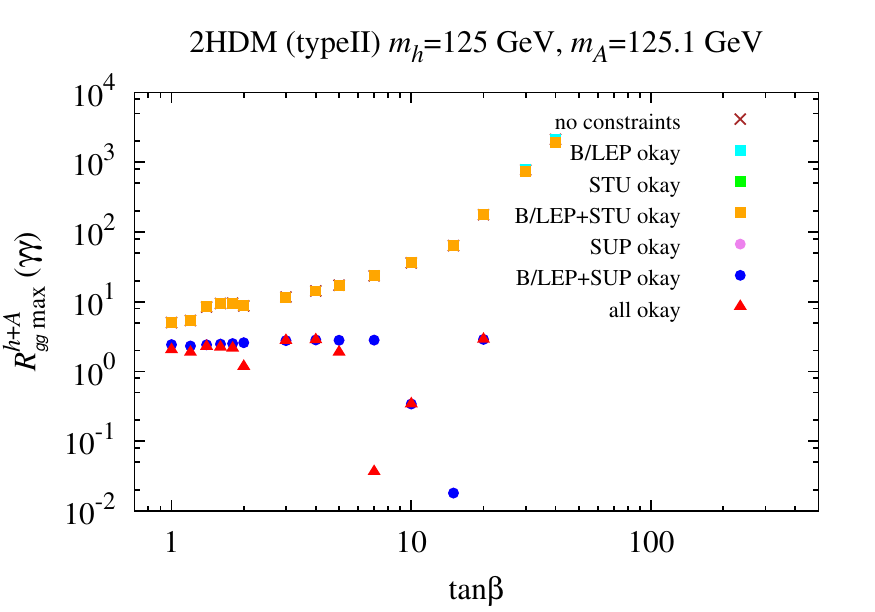}
\end{center}
\vspace{-.2in}
\caption{maximal $R^{h+A}_{gg}(\gam\gam)$  vs. $\tanb$ when $m_h=125\gev,m_A=125.1\gev$ after imposing various 
constraints.}
\label{hAonly}
\end{figure}

The signal at $125\gev$ cannot be pure pseudoscalar Higgs $A$ since it does not couple to $ZZ$, a final state that is definitely 
present at $125\gev$.  However, one can imagine that the scalar Higgs $h$ or $H$ and the $A$ both have mass close to $125\gev$ 
and that the net $\gam\gam$ rate gets substantial contributions from both the $h$ and the $A$ while only the former contributes to the $ZZ$ rate. This possibility is explored in Fig.~\ref{hAonly}.

For the Type I model, $R^h_{gg}(\gam\gam)$ is significantly enhanced only for the same $\tanb=3, 4\,\text{and}\,20$
values as in the case of having (only) $m_h=125\gev$ and the pseudoscalar contribution  $R^A_{gg}(\gam\gam)$ turns out to be tiny.
However, the contribution to the  $\tau\tau$ final state from the $A$ {\it can} be substantial if $\tanb$ is small since the coupling $A$ to leptons is proportional to $\cot\beta$. In the end, only $\tan\beta=20$  yields both an enhanced $\gam\gam$ rate, $R_{gg\, {\rm max}}^{h+A}(\gam\gam)=1.31$, and SM-like rates for the $ZZ$ and $\tau\tau$ final states.

For the Type II model, the enhancement of $R_{gg}^{h+A}(\gam\gam)$ is essentially the same as that for $R_{gg}^h(\gam\gam)$ for the case when only $m_h=125\gev$, reaching maximum values of order $2-3$. However, as in the pure $m_h=125\gev$ case, a substantial enhancement  of $R_{gg}^{h+A}(\gam\gam)$ is most often associated with $R_{gg}^{h+A}(ZZ)>R_{gg}^{h+A}(\gam\gam)$ (contrary to the center values by the current LHC data). But this is not {\it always} the case. The exception occurs at $\tanb\sim1$, at which there exist theoretically consistent parameter choices for which $R^{h+A}_{gg}(\gam\gam) \sim R^{h+A}_{gg}(ZZ) \sim 1.6$, fully consistent with the Moriond 2013 ATLAS results. But unfortunately for those points $\tau\tau$ signal is predicted to be too strong, $R^{h+A}_{gg}(\tau\tau) > 3.75$, a value far above than what is observed at the CMS analysis.

\section{Conclusions}

We have analyzed the Type I and Type II two-Higgs-doublet models with regard to consistency with a significant enhancement of the gluon fusion induced $\gamma\gamma$ signal at the LHC at $\sim 125\gev$, as seen in the ATLAS data set, but possibly not in the CMS results presented at Moriond 2013. All possible theoretical and experimental constraints have been imposed. We find that the conditions coming from requiring vacuum stability, unitarity and perturbativity play the key role in limiting the maximal possible enhancement. Generically, we conclude that the Type II model allows a maximal enhancement of order of $2-3$, whereas within the Type I model the maximal enhancement is limited to  $\lsim 1.3$. 

Moriond 2013 ATLAS results suggest an enhancement for gluon fusion induced $\gam\gam$ rate of order $1.6$ (but with large errors). Only Type II models can give such a large value. However,  we find  that in the Type II model the parameters that give $R^h_{gg}(\gam\gam)\sim 1.6$ are characterized by $R^h_{gg}(ZZ)\sim 1.5 R^h_{gg}(\gam\gam)$, a result that is  inconsistent with the ATLAS central value of $R^h_{gg}(ZZ) \sim 1.5$.  Similar statements apply to the case of the heavier $H$ having a mass of $125\gev$.  Furthermore, under approximately degenerate $h$ and $A$ at $125\gev$ scenario, although observed center values $\sim 1.5-1.6$ in both $\gam\gam$ and $ZZ$ channels could be simultaneously satisfied, it suffers from an unwanted too large $\tau\tau$ rate. In short, Type II models are unable to give a significantly enhanced gluon fusion induced $\gam\gam$ signal while maintaining consistency with other channels. In contrast, the CMS data suggests values of $R^h_{gg}(\gam\gam)<1$ and  $R^h_{gg}(ZZ)\sim 1$, as easily obtained in the Type II model context.

Once again, we emphasize that a mild enhancement of the $\gam\gam$ rate, while other final states, in particular $ZZ$,  have close to SM rates, {\it is} possible for either single $h$ or $h$ and $A$ degeneracy having mass at $125 \gev$ in Type I models. For these scenarios, the charged Higgs is light, $m_{H^\pm}=90\gev$, which does not conflict with LHC data due to small $BR(t\to H^+b)$. Thus, Type I models could provide a consistent picture if the LHC results converge to only a modest enhancement for $R^h_{gg}(\gam\gam)\lsim 1.3$.

\section*{Acknowledgments}

The attendance and presentation at the conference was supported in part by US DOE grant DE-FG03-91ER40674. YJ thanks the supervision of JFG and collaboration with BG and AD. YJ also thank the organizers of the Moriond EW conference for the kind invitation and for the generous grant award to cover the local living expense for a young scientist.


\section*{References}

\begin{thebibliography}{99}
\bibitem{Drozd:2012vf}
  A.~Drozd, B.~Grzadkowski, J.~F.~Gunion and Y.~Jiang,
  arXiv:1211.3580 [hep-ph].
 
\bibitem{Aad:2012tfa}
  G.~Aad {\it et al.}  [ATLAS Collaboration],
  Phys.\ Lett.\ B {\bf 716} (2012) 1
  [arXiv:1207.7214 [hep-ex]].
  
\bibitem{Chatrchyan:2012ufa}
  S.~Chatrchyan {\it et al.}  [CMS Collaboration],
  Phys.\ Lett.\ B {\bf 716} (2012) 30
  [arXiv:1207.7235 [hep-ex]].   

\bibitem{Aaltonen:2012qt}
  T.~Aaltonen {\it et al.}  [CDF and D0 Collaborations],
  Phys.\ Rev.\ Lett.\  {\bf 109} (2012) 071804
  [arXiv:1207.6436 [hep-ex]].
  
\bibitem{ATLAS:2012klq}
  [ATLAS Collaboration],
  ATLAS-CONF-2012-170.
  
\bibitem{Chatrchyan:2013lba}
  S.~Chatrchyan {\it et al.}  [CMS Collaboration],
  arXiv:1303.4571 [hep-ex].
  
\bibitem{ATLAS:2013oma}
  [ATLAS Collaboration],
  ATLAS-CONF-2013-012.
   
\bibitem{ATLAS:2013nma}
  [ATLAS Collaboration],
  ATLAS-CONF-2013-013.
  
\bibitem{CMS:ril}
  [CMS Collaboration],
  CMS-PAS-HIG-13-001.
  
\bibitem{CMS:xwa}
  [CMS Collaboration],
  CMS-PAS-HIG-13-002.

\bibitem{Gunion:2012gc}
  J.~F.~Gunion, Y.~Jiang and S.~Kraml,
  Phys.\ Rev.\ D {\bf 86} (2012) 071702
  [arXiv:1207.1545 [hep-ph]].
  
\end{thebibliography}
\end{document}